\documentclass{jps-cp}
\usepackage{txfonts} 
\usepackage{bm}

\title{Conversion between the formal and observed parameters in $R$-matrix theory}

\author{Masahiko \textsc{Katsuma}$^{1,2}$}

\inst{
  $^{1}$Advanced Mathematical Institute, Osaka City University, Osaka 558-8585, Japan \\
  $^{2}$Institut d'Astronomie et d'Astrophysique, Universit\'e Libre de Bruxelles, B-1050 Brussels, Belgium}

\email{mkatsuma@sci.osaka-cu.ac.jp; mkatsuma@gmail.com}

\recdate{July 19, 2019}

\abst{
  The conversion between the formal and observed parameters is obtained from the phase equivalence between the $R$-matrix and Breit-Wigner formula, and it is applied to a study of the low-energy $E$1 $S$-factor of $^{12}$C($\alpha$,$\gamma_0$)$^{16}$O.
  As an example, weak interference between 1$^-$ states in $^{16}$O is discussed.
  As well as the ordinary $R$-matrix method, the present method calculates the $S$-matrix and cross sections, independent of the boundary condition.
  Therefore, the $E$1 $S$-factor at $E_{c.m.} = 300$ keV is found to be reduced from the current evaluation.
}

\kword{$R$-matrix method, Resonance parameters, $^{12}$C($\alpha$,$\gamma$)$^{16}$O}

\begin{document}
\maketitle

\section{Introduction}
  Low-energy reactions are usually thought to be governed by the complicated process of nucleus reactions.
  So, the phenomenological $R$-matrix method \cite{Lan58,Des} can be applied to evaluate the experimental cross section data.
  In the evaluation, it should be noted that the resonance parameters in the $R$-matrix are different from the observed experimental values.

  In XII-3 of \cite{Lan58}, the {\it observed} parameters are defined as the experimental quantities of Breit-Wigner form, and the {\it formal} parameters are the model parameters in $R$-matrix theory.
  The conversion between the formal and observed parameters is given by the phase ($S$-matrix) equivalence between the $R$-matrix and Breit-Wigner formula, and it ensures that the calculations correspond to the experimental width and cross sections.
  If the reduced width is narrow, the formal parameters are almost equivalent to the observed experimental width and resonance energy.
  However, the difference might have to be considered accurately when the reduced width is a value close to the single-particle limit which represents the well-developed cluster state, e.g. 1$^-_2$ state ($E_x= 9.59$ MeV) in $^{16}$O \cite{Kat10,Fuj80}.

  In the present report, I review the conversion between the formal and observed parameters, and I apply it to a study of the $E$1 cross sections for $^{12}$C($\alpha$,$\gamma_0$)$^{16}$O.
  As an example, I discuss the weak interference between the 1$^-$ states in $^{16}$O \cite{Kat17}.
  After the conversion, the $S$-matrix and cross sections can be calculated independent of the boundary condition parameter \cite{Bar72}.
  With any choice of the boundary condition, I show that the $E$1 $S$-factor at $E_{c.m.}=300$ keV is reduced from the current evaluation.

\section{Resonance parameters}

  In $R$-matrix theory \cite{Lan58,Des}, the $S$-matrix is given by
  \begin{eqnarray}
    S_L &=& \frac{I_L(k a_c)}{O_L(k a_c)} \cdot
    \frac{1-[L_L^\ast(k a_c)-a_c \tilde{b}_c] R_L}{1-[L_L(k a_c)-a_c \tilde{b}_c]R_L}
    \hspace{2mm}=\hspace{2mm} \exp\left[\,2i(\delta^{\rm HS}_L+\delta^{\rm R}_L)\,\right],
    \label{eq:smat}
  \end{eqnarray}
  where $I_L(k r)$ and $O_L(k r)$ are the incoming and outgoing Coulomb wave functions, respectively.
  $L$ is the angular momentum of the relative motion between nuclei in the channel region.
  $k$ is the wavenumber $k=\sqrt{2\mu|E_{c.m.}|/\hbar}$; $\mu$ is the reduced mass; $E_{c.m.}$ is the center-of-mass energy.
  $a_c$ is the channel radius.
  $\tilde{b}_c$ is the boundary condition parameter, the logarithmic derivative of the internal wavefunction $\tilde{\varphi}_{nL}$, $\tilde{b}_c=d \ln\tilde{\varphi}_{nL}(a_c)/dr$.
  $L_L(k a_c)$ is the logarithmic derivative of the outgoing Coulomb function.
  The real and imaginary parts of $L_L(k a_c)$ are the shift function $\Delta_L(E_{c.m.}, a_c)$ and the penetration factor $P_L(E_{c.m.}, a_c)$, respectively.
  $R_L$ denotes the $R$-matrix,
  \begin{eqnarray}
    R_L(E_{c.m.}) &=& \mathop{\sum}_{n} \frac{\tilde{\gamma}_{n L}^2}{\tilde{E}_{n L}-E_{c.m.}} + {\cal R}_{\alpha L},
    \label{eq:rmat}
  \end{eqnarray}
  where $\tilde{E}_{n L}$ and $\tilde{\gamma}_{n L}^2$ are the {\it formal} resonance energy and {\it formal} reduced width, respectively.
  ${\cal R}_{\alpha L}$ denotes an energy-independent component.
  $n$ is the ordinal number of the state with $L$ in order of the excitation energy.
  $\delta^{HS}_L$ in Eq.~(\ref{eq:smat}) is the hard-sphere phase shift, and $\delta^{R}_L$ is defined by
  \begin{eqnarray}
    \delta^{\rm R}_L &=& \arctan\!\left( \frac{P_L(E_{c.m.}, a_c)R_L(E_{c.m.})}{1-\Delta_L^{(b)}(E_{c.m.}, a_c)R_L(E_{c.m.})}\right),
    \label{eq:ps-r}
  \end{eqnarray}
  where $\Delta_L^{(b)}(E_{c.m.}, a_c)\equiv\Delta_L(E_{c.m.}, a_c)-a_c \tilde{b}_c$.
  For a Breit-Wigner resonance, the $S$-matrix is given as
  \begin{eqnarray}
    S_L &=& e^{2i\delta^{\rm bg}} \frac{E_{c.m.}-E_{nL}-i\Gamma_{nL}/2}{E_{c.m.}-E_{nL}+i\Gamma_{nL}/2}
    \hspace{2mm}=\hspace{2mm}\exp\left[\,2i(\delta^{\rm bg}+\delta^{\rm BW})\,\right],
    \label{eq:smat-bw}
  \end{eqnarray}
  where $E_{nL}$ and $\Gamma_{nL}$ are the {\it observed} resonance energy and width, $\Gamma_{nL}=2P_L(E_{c.m.},a_c)\gamma^2_{nL}$; $\gamma^2_{nL}$ is the {\it observed} reduced width.
  $\delta^{\rm bg}$ is the background phase shift, and $\delta^{\rm BW}_L$ is defined by
  \begin{eqnarray}
    \delta^{\rm BW}_L &=& \arctan\left( \frac{\Gamma_{nL}/2}{E_{nL}-E_{c.m.}}\right).
    \label{eq:ps-bw}
  \end{eqnarray}

  The conversion between the formal and observed parameters is obtained in single-pole from Eqs.~(\ref{eq:ps-r}) and (\ref{eq:ps-bw}).
  So, the formal reduced width is given as
  \begin{eqnarray}
    \tilde{\gamma}_{n L}^2(E_{c.m.}) &=& 
    \frac{\gamma_{n L}^2}{1-\gamma_{n L}^2 \Delta_L^\prime(E_{n L}, a_c)
    \left[\, 1+Q_{n L}(E_{c.m.}, a_c)\,\right]},
    \label{eq:g-ecm}
  \end{eqnarray}
  where $\Delta_L^\prime=d\Delta_L/dE_{c.m.}$.
  $Q_{nL}$ is a higher-order term of the linear approximation to $\Delta_L(E_{c.m.},a_c)$ \cite{Kat17}.
  For most of reactions, $Q_{nL}=0$ is widely used.  
  The formal energy $\tilde{E}_{n L}$ is defined in
  \begin{eqnarray}
    \tilde{E}_{n L}(E_{c.m.}) &=& E_{n L} + \tilde{\gamma}_{n L}^2(E_{c.m.}) \Delta_L^{(b)}(E_{n L}, a_c)\,(\,1+d_{n L}\,),
    \label{eq:er-ecm}
  \end{eqnarray}
  where $d_{n L}$ is a parameter stemming from multi-levels in Eq.~(\ref{eq:rmat}).
  $d_{n L}$ and the observed parameters \cite{Til93} are adjusted self-consistently so as to satisfy the relation of
  \begin{eqnarray}
    \Delta_L^{(b)}(E_{nL},a_c)\, R_L(E_{nL}) &=& 1.
    \label{eq:SR}
  \end{eqnarray}
  The values of the parameters are listed in \cite{Kat17}.
  If the reduced width is narrow, the formal parameters are found to be almost identical to the observed ones, $\tilde{\gamma}_{n L}^2 \approx \gamma_{n L}^2$ and $\tilde{E}_{nL}\approx E_{nL}$ in Eqs.~(\ref{eq:g-ecm}) and (\ref{eq:er-ecm}).
  
  The $S$-matrix of Eq.~(\ref{eq:smat}) and all cross sections can be calculated independent of $\tilde{b}_c$.
  To review this requirement, let me recall the transformation of the formal parameter set ($\tilde{b}_c, \tilde{E}_{nL}, \tilde{\gamma}_{nL}$) \cite{Bar72}.
  The mathematically-transformed set ($\tilde{b}_c^\prime, \tilde{E}_{nL}^\prime, \tilde{\gamma}_{nL}^\prime$) is obtained from diagonalization of a matrix given as
  \begin{eqnarray}
    \bm{K}\bm{C}\bm{K^t}=\bm{D},
    \label{eq:diag}
  \end{eqnarray}
  where $D_{ij}=\tilde{E}_{iL}^\prime\delta_{ij}$.
  $\bm{K}$ is an orthogonal matrix.
  $\bm{C}$ is defined in $C_{ij} = \tilde{E}_{iL}\delta_{ij} -a_c(\tilde{b}_c^\prime -\tilde{b}_c)\tilde{\gamma}_{iL}\tilde{\gamma}_{jL}$.
  The transformed reduced width amplitudes are obtained from $\tilde{\gamma}_{iL}^\prime = \mathop{\sum}_j K_{ij} \tilde{\gamma}_{jL}$.
  In other words, Eq.~(\ref{eq:diag}) is the transformation between two formal parameter sets under invariance of the $S$-matrix in Eq.~(\ref{eq:smat}).
  It should be noted that the present method is the same as the ordinary method after the conversion of Eqs.~(\ref{eq:g-ecm}) and (\ref{eq:er-ecm}).

\section{Results}

\begin{figure}[t]
    \vspace{10mm}
    \begin{center}
      \begin{tabular}{cc}
        \includegraphics[width=0.385\linewidth]{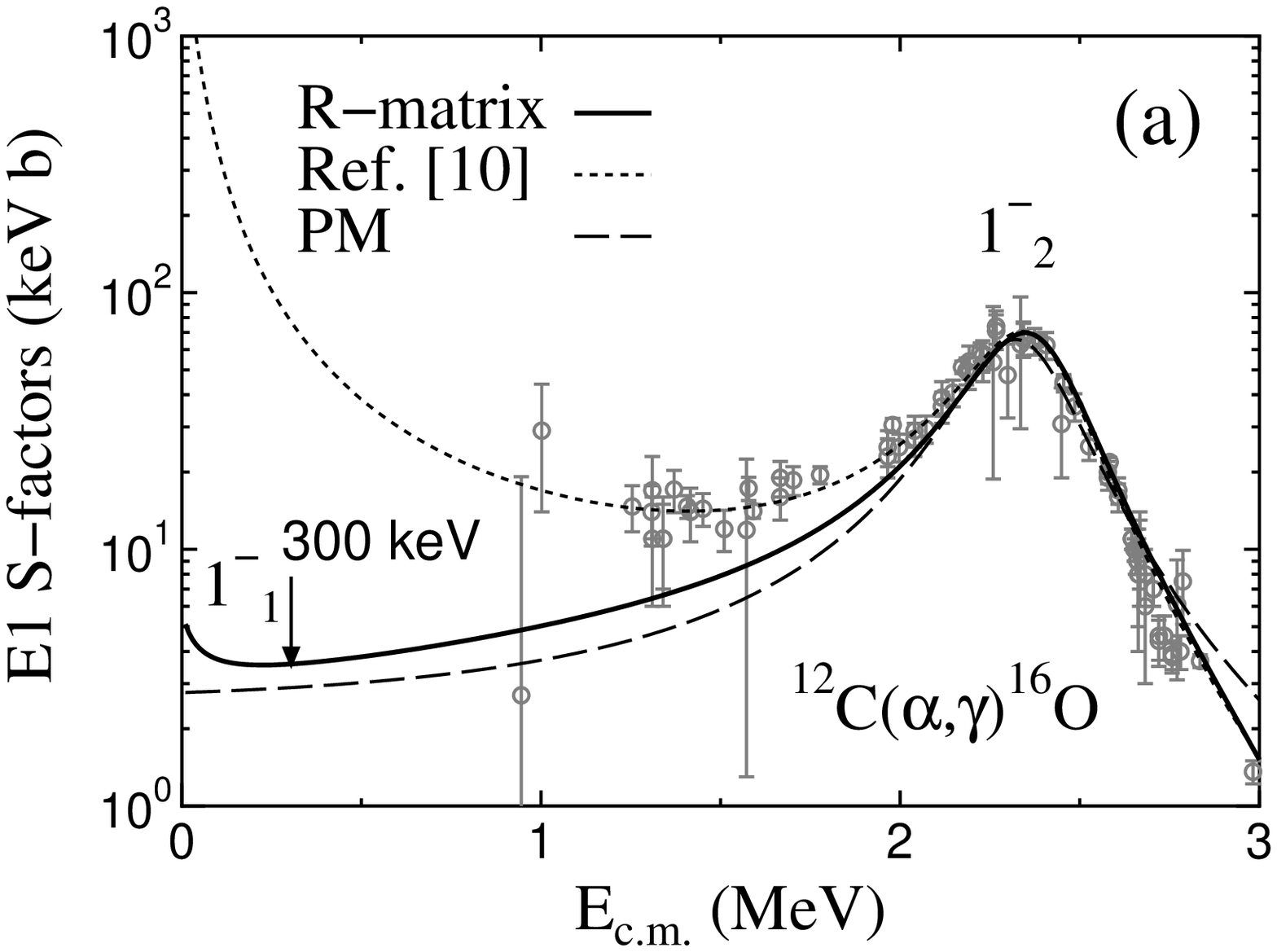} &
        \includegraphics[width=0.38\linewidth]{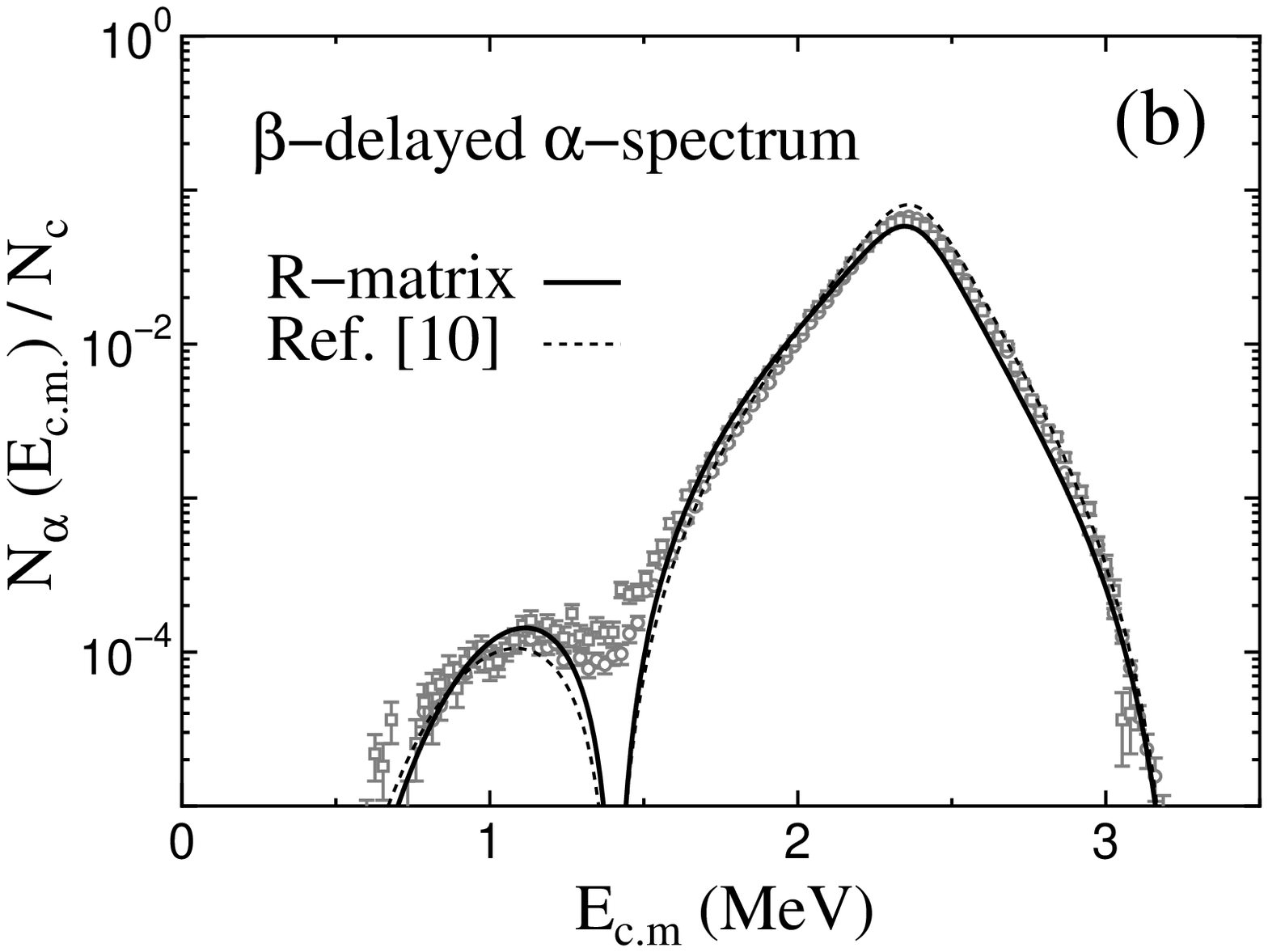}
      \end{tabular}
    \end{center}
    \vspace{-26mm}
    \caption{
      (a) $E$1 $S$-factor for $^{12}$C($\alpha$,$\gamma_0$)$^{16}$O,
      (b) $\beta$-delayed $\alpha$-particle spectrum of $^{16}$N.
      The solid curves are the present example \cite{Kat17}.
      The dotted and dashed curves are the results of the $R$-matrix method \cite{Azu94} and the potential model \cite{Kat08}, respectively.
      The experimental data are taken from \cite{Kun01} and \cite{Azu94,Tan10}.
    }
    \label{fig:sfact}
  \end{figure}

  In Fig.~\ref{fig:sfact}(a), I show the calculated $E$1 $S$-factors for $^{12}$C($\alpha$,$\gamma_0$)$^{16}$O.
  The solid curve is the present example, including 1$^-_1$ ($E_x= 7.12$ MeV), 1$^-_2$, 1$^-_3$ ($E_x= 12.44$ MeV), and 1$^-_4$ ($E_x= 13.09$ MeV) of $^{16}$O \cite{Til93}.
  In the present example, the interference between 1$^-_1$ and 1$^-_2$ appears to be weak, and the resultant $E$1 $S$-factor is reduced from the current evaluation at $E_{c.m.}=300$ keV corresponding to the Helium burning temperature.
  The dotted and dashed curves are the results of the $R$-matrix method \cite{Azu94} and the potential model \cite{Kat08}, respectively.
  The fit to the $S$-factor data \cite{Kun01} looks good for the dotted curve.
  However, it fails to reproduce the $\alpha$-width of 1$^-_2$, $\Gamma_{21}=359$ keV \cite{Azu94}.
  {\it cf.} $\Gamma_{21}= (420\pm20)$ keV \cite{Til93} and $\Gamma_{21}= 432$ keV (the present example).
  It should be noted that the reduced width is assumed to be narrow in \cite{Azu94}.
  The present example shown by the solid curve in Fig.~\ref{fig:sfact}(b) reproduces the experimental $\beta$-delayed $\alpha$-spectrum of $^{16}$N in the same quality of the fits as \cite{Azu94} (dotted curve).
  In other words, the small $E$1 $S$-factor at low energies is in agreement with the experimental data of $\beta$-delayed $\alpha$-spectrum of $^{16}$N \cite{Azu94,Tan10}.
  The present example is also consistent with the experimental result of $p$-wave phase shifts of $\alpha$+$^{12}$C elastic scattering \cite{Kat17}.

\begin{figure}[t]
    \vspace{41mm}
    \centerline{\includegraphics[width=0.965\linewidth]{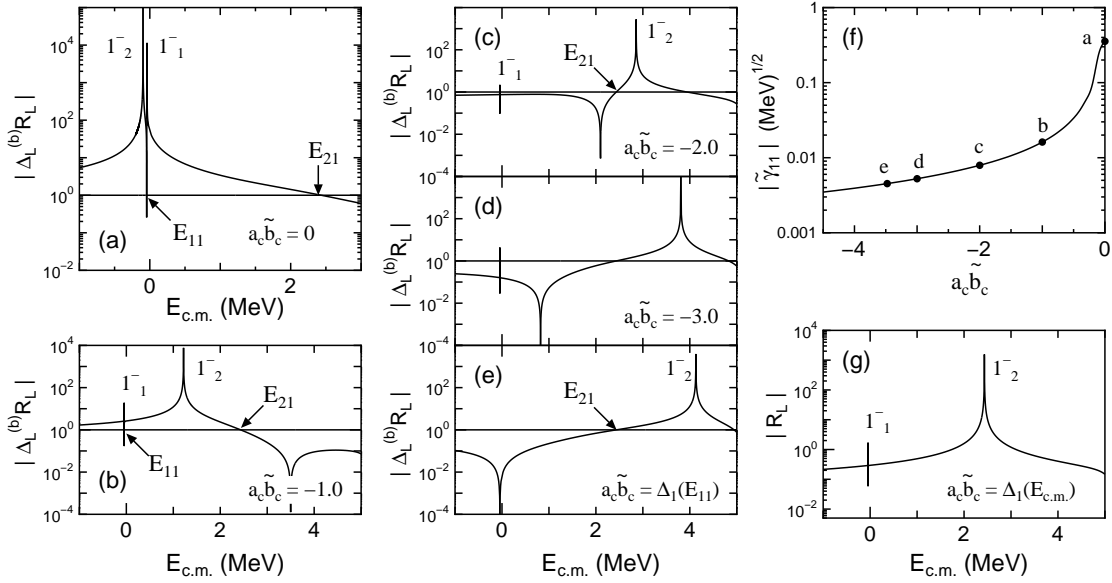}}
    \vspace{-53mm}
    \caption{
      (a)-(e) $R$-matrix multiplied by $\Delta^{(b)}_1$.
      The solid curves are obtained from the present $R$-matrix method.
      $a_c\tilde{b}_c= 0$, -1, -2, -3 and $\Delta_1(E_{11},a_c)$ are used.
      (f) Reduced width amplitudes of 1$^-_1$ as a function of $a_c\tilde{b}_c$.
      The solid circles indicate the used values in (a)-(e).
      (g) Derived $R$-matrix with $a_c\tilde{b}_c=\Delta_1(E_{c.m.},a_c)$.
      The peaks are the energy positions of $E_{n1}=\tilde{E}_{n1}$.
      The displayed $R$-matrices make the same cross sections as those in Fig.~\ref{fig:sfact}.
    }
    \label{fig:bc}
  \end{figure}

  The corresponding $R$-matrix multiplied by $\Delta_1(E_{c.m.},a_c)$ is illustrated in Fig.~\ref{fig:bc}(a), $\tilde{b}_c=0.$
  The peak is the formal energy and the energy position of Eq.~(\ref{eq:SR}) is the observed resonance energy.
  The 1$^-_2$ peak is shifted below $E_{c.m.}=0$ close to the pole of 1$^-_1$, because of the large reduced width.
  The proximity of two poles appears to make their interference weak in the present calculation.
  In addition, the contribution from the 1$^-_2$ state seems to dominate the magnitude of the $R$-matrix below the barrier.

  After the conversion of Eqs.~(\ref{eq:g-ecm}) and (\ref{eq:er-ecm}), the transformation of Eq.~(\ref{eq:diag}) makes the same $S$-factors as in Fig.~\ref{fig:sfact}, because the $S$-matrix is invariant.
  So, the feature of weak interference can be also seen in other sets of the formal resonance parameter.
  Figure \ref{fig:bc}(b)--\ref{fig:bc}(e) shows the $R$-matrix multiplied by $\Delta_1^{(b)}$ at $a_c\tilde{b}_c =-1$, -2, -3, and $\Delta_1(E_{11},a_c)$.
  As $a_c \tilde{b}_c$ decreases, the peak of the 1$^-_2$ state is shifted to higher energies, whereas the energy position of Eq.~(\ref{eq:SR}) remains on $E_{21} = 2.434$ MeV.
  In contrast, the energy position of 1$^-_1$ does not move, and the absolute value of $\tilde{\gamma}_{11}$ becomes small for $a_c\tilde{b}_c < 0$. (Fig.~\ref{fig:bc}(f))
  Consequently, the 1$^-_1$ state is isolated, and it does not appear to interfere with other states.

  If $a_c\tilde{b}_c= \Delta_1(E_{c.m.},a_c)$ is used, the formal parameters are equal to the observed parameters.
  This type of definition is used in \cite{Azu94,Bru02}.
  The derived $R$-matrix is shown in Fig.~\ref{fig:bc}(g).
  The peak of 1$^-_2$ can be seen at $E_{c.m.}=2.434$ MeV, and the very sharp peak can be found at $E_{c.m.}=-0.0451$ MeV.
  The resulting $S$-factors and $\beta$-delayed $\alpha$-spectrum are the same as those in Fig.~\ref{fig:sfact}.

\section{Summary}
  The conversion between the formal and observed resonance parameters has been obtained from the phase equivalence between the $R$-matrix and Breit-Wigner formula, and it has been applied to the $E$1 $S$-factor of $^{12}$C($\alpha$,$\gamma_0$)$^{16}$O.
  Using the present $R$-matrix method, I have discussed the weak interference between 1$^-_1$ and 1$^-_2$ and the resulting small $E$1 $S$-factor at $E_{c.m.}= 300$ keV.
  In the calculation, the pole energy of 1$^-_2$ is found to be located in the vicinity of 1$^-_1$.
  This proximity suppresses their interference, and it consequently makes the small $E$1 $S$-factor below the barrier.
  This is caused by the accurate treatment of the broad resonance of 1$^-_2$.
  The present example is consistent with the experimental results of $\alpha$-particle decay width of 1$^-_2$, $\beta$-delayed $\alpha$-particle spectrum of $^{16}$N, and $p$-wave phase shift of $\alpha$+$^{12}$C elastic scattering.
  The formal parameter set is transformed into other formal sets, without changing the $S$-matrix, i.e. all cross sections.
  So, the weak interference can be seen even when other sets are used.
  Therefore, the $E$1 $S$-factor is found to be reduced from the enhanced value of the current evaluation, by the $R$-matrix method independent of the boundary condition.

  I thank M.~Arnould, A.~Jorissen, K.~Takahashi, H.~Utsunomiya, Y.~Ohnita, and Y.~Sakuragi for their hospitality during his stay at Universit\'e Libre de Bruxelles and Osaka City University.


\begin{thebibliography}{99}
\bibitem{Lan58}
  A.~M. Lane and R.~G. Thomas, Rev. Mod. Phys. {\bf 30}, 257 (1958).
\bibitem{Des}
  P. Descouvemont, {\it Theoretical Models for Nuclear Astrophysics} (Nova Science, NY, 2003);
  I.~J. Thompson and F. Nunes, {\it Nuclear Reactions for Astrophysics} (Cambridge Univ. Press, NY, 2009).
\bibitem{Kat10}
  M. Katsuma, Phys. Rev. C {\bf 81}, 067603 (2010); J. Phys. G {\bf 40}, 025107 (2013); EPJ Web of Conf. {\bf 66}, 03041 (2014).
\bibitem{Fuj80}
  Y.~Fujiwara et al., Prog. Theor. Phys. Suppl. {\bf 68}, 29 (1980).
\bibitem{Kat17}
  M. Katsuma, {\it NIC XV}, Springer Proceedings in Physics {\bf 219}, 389 (2019); arXiv:1701.02848. 
\bibitem{Bar72}
  F.~C. Barker, Aust. J. Phys. {\bf 25}, 341 (1972).
\bibitem{Til93}
  D.~R. Tilley, H.~R. Weller and C.~M. Cheves, Nucl. Phys. A {\bf 564}, 1 (1993).
\bibitem{Kat08}
  M. Katsuma, Phys. Rev. C {\bf 78}, 034606 (2008); ibid. C {\bf 81}, 029804 (2010); Astrophys. J. {\bf 745}, 192 (2012); Phys. Rev. C {\bf 90}, 068801 (2014); JPS Conf. Proc. {\bf 14}, 021009 (2017).
\bibitem{Kun01}
  R. Kunz et al., Phys. Rev. Lett. {\bf 86}, 3244 (2001);
  M. Assun\c c\~ao et al., Phys. Rev. C {\bf 73}, 055801 (2006);
  H. Makii et al., ibid. {\bf 80}, 065802 (2009);
  R. Plag et al., ibid. {\bf 86}, 015805 (2012);
  J.~M.~L. Ouellet et~al., ibid. {\bf 54}, 1982 (1996).
\bibitem{Azu94}
  R.~E. Azuma et al., Phys. Rev. C {\bf 50}, 1194 (1994).
\bibitem{Tan10}
  X.~D. Tang et al., Phys. Rev. C {\bf 81}, 045809 (2010).
\bibitem{Bru02}
  C.~R. Brune et al., Phys. Rev. C {\bf 66}, 044611 (2002).
\end{thebibliography}
\end{document}